\documentclass[runningheads]{llncs}
\usepackage[utf8]{inputenc}
\usepackage{cite}
\usepackage{amsmath, amssymb}
\usepackage{physics}
\usepackage{xcolor}
\usepackage{mathtools}
\usepackage[shortlabels]{enumitem}
\usepackage[colorlinks=true,citecolor=blue,linkcolor=blue,urlcolor=blue,breaklinks]{hyperref}
\usepackage[nolist]{acronym}
\usepackage{mathdots}
\usepackage{centernot}
\usepackage{bm}
\usepackage{graphicx}

\graphicspath{{figures/}}

\newcommand{\PTr}[2]{\text{Tr}_{#1}\left\{#2\right\}}

\newcommand{\cF}{\mathcal{F}}
\newcommand{\cZ}{\mathcal{Z}}
\newcommand{\cS}{\mathcal{S}}

\newcommand{\cO}{\mathcal{O}}
\newcommand{\cH}{\mathcal{H}}

\newcommand{\cP}{\mathcal{P}}

\newcommand{\vZ}{\bm{Z}}
\newcommand{\vU}{\bm{U}}
\newcommand{\vH}{\bm{H}}
\newcommand{\vV}{\bm{V}}
\newcommand{\vE}{\bm{E}}
\renewcommand{\vb}{\bm{b}}
\newcommand{\vw}{\bm{w}}

\begin{document}

\title{Variational Quantum Algorithms for Gibbs State Preparation}

\author{Mirko Consiglio \orcidID{0000-0001-8730-0206}}

\authorrunning{M. Consiglio}

\institute{Department of Physics, University of Malta, Msida MSD 2080, Malta \email{mirko.consiglio@um.edu.mt}}

\maketitle

\emergencystretch 3em

\begin{abstract}
    Preparing the Gibbs state of an interacting quantum many-body system on \ac{NISQ} devices is a crucial task for exploring the thermodynamic properties in the quantum regime. It encompasses understanding protocols such as thermalization and out-of-equilibrium thermodynamics, as well as sampling from faithfully prepared Gibbs states could pave the way to providing useful resources for quantum algorithms. \Acp{VQA} show the most promise in efficiently preparing Gibbs states, however, there are many different approaches that could be applied to effectively determine and prepare Gibbs states on a \ac{NISQ} computer. In this paper, we provide a concise overview of the algorithms capable of preparing Gibbs states, including joint Hamiltonian evolution of a system--environment coupling, quantum imaginary time evolution, and modern \acp{VQA} utilizing the Helmholtz free energy as a cost function, among others. Furthermore, we perform a benchmark of one of the latest variational Gibbs state preparation algorithms, developed by Consiglio et al.~\cite{Consiglio2023}, by applying it to the spin 1/2 one-dimensional $XY$ model.
    

\end{abstract}

\begin{acronym}
\acro{PQC}{parametrized quantum circuit}
\acro{VQA}{variational quantum algorithm}
\acro{NISQ}{noisy intermediate-scale quantum}
\acro{BFGS}{Broyden--Fletcher--Goldfarb--Shanno}
\acro{SPSA}{simultaneous perturbation stochastic approximation}
\acro{TFD}{thermofield double}
\acro{QAOA}{quantum approximate optimization algorithm}
\acro{QITE}{quantum imaginary time evolution}
\acro{METTS}{minimally entangled typical thermal states}
\acro{ADAPT}{adaptive derivative assembled problem-tailored}
\acro{QBM}{quantum Boltzmann machine}
\acro{CBM}{classical Boltzmann machine}
\end{acronym}

\section{Introduction}
\acresetall

Gibbs states (also known as thermal states) can be used for quantum simulation~\cite{Childs2018}, quantum machine learning~\cite{Kieferova2017, Biamonte2017}, quantum optimization~\cite{Somma2008}, and the study of open quantum systems~\cite{Poulin2009}. Moreover, semi-definite 
programming~\cite{Brandao2016}, combinatorial optimization problems~\cite{Somma2008}, and training quantum Boltzmann machines~\cite{Kieferova2017}, can be tackled by sampling from well-prepared Gibbs states. Nevertheless, the preparation of an arbitrary quantum state on a quantum computer, is a \textsf{QMA}-hard problem~\cite{Watrous2008}. Preparing Gibbs states, specifically at low temperatures, could be as hard as finding the ground-state of that Hamiltonian~\cite{Aharonov2013}.

From a physical point of view, A Gibbs state is the quantum state that is at thermal equilibrium with the surrounding environment, in the canonical ensemble. Let's consider a Hamiltonian $\cH$, describing $n$ interacting qubits. The Gibbs state at inverse temperature $\beta \equiv 1 / (k_\text{B} T)$, where $k_\text{B}$ Boltzmann constant and $T$ is the temperature, is defined as
\begin{equation}
    \rho(\beta, \cH) = \frac{e^{-\beta \cH}}{\cZ(\beta, \cH)},
\end{equation}
where the partition function $\cZ(\beta, \cH)$ is
\begin{equation}
    \cZ(\beta, \cH) = \Tr{e^{-\beta \cH}} = \sum_{i=0}^{d - 1} e^{-\beta E_i}.
\end{equation}
Here the dimension $d = 2^n$, while $\{E_i\}$ are the eigenenergies of $\cH$ (with $\{\ket{E_i}\}$ denoting the corresponding eigenstates), i.e. $\cH\ket{E_i} = E_i\ket{E_i}$.

While it may be a straightforward procedure to allow a system to naturally thermalize with its environment, to obtain a Gibbs state, there is a great significance in being able to obtain a Gibbs of any arbitrary Hamiltonian. Since most physical systems are bound to specific types of Hamiltonian, a quantum computer can --- at least in principle --- simulate any arbitrary quantum system. Moreover, while a Gibbs distribution can be prepared on a classical computer, or even have a density matrix describing the Gibbs state stored on it, is not equivalent to having an actual Gibbs state on a quantum computer. Since we can continue to evolve, probe and simulate the Gibbs state to study quantum systems. As Richard Feynman aptly said~\cite{Trabesinger2012}, we require a quantum system to be able to effectively simulate, another, quantum system.

The objective of this study is twofold: firstly, to offer a comprehensive overview of Gibbs state preparation algorithms in Section~\ref{sec:overview}, and to explore a noteworthy application of Gibbs states in quantum Boltzmann machines. Secondly, to present the recent variational Gibbs state preparation algorithm by Consiglio et al.~\cite{Consiglio2023} in Section~\ref{sec:VGSP}. In Section~\ref{sec:results}, we apply the \ac{VQA} to prepare Gibbs states of the $XY$ model, while also qualitatively investigating the scalability of the algorithm. Finally in Section.~\ref{sec:conclusion}, we draw our conclusions.

\section{Overview of Gibbs State Preparation Algorithms} \label{sec:overview}

The first computational technique one would usually resort to prepare a Gibbs state would be exact diagonalisation~\cite{Jaklic1994}, however, since the full energy spectrum is needed, this method quickly becomes infeasible once the system size grows. A different approach is based on thermal pure quantum states~\cite{Sugiura2012, Sugiura2013}, which significantly reduces computational overhead when compared with exact diagonalisation. Nevertheless, since it requires preparing and evolving random quantum states, it is still limited in terms of system size. On the other hand, a quantum algorithm based on evaluating thermal pure shadows that operates using a quantum computer was proposed~\cite{Coopmans2023}.

Initially, the first algorithms for preparing Gibbs states on a quantum computer were based on the idea of coupling the system to a register of ancillary qubits, and letting the system and environment evolve under a joint Hamiltonian, simulating the physical process of thermalization~\cite{Terhal2000, Shabani2016, Metcalf2020}. Davies~\cite{Davies1974, Davies1976} showed that, in the limits of small time-steps or weak system-bath interactions, the dynamics are described by a Lindblad master equation --- the Davies generator --- that thermalizes the system in the long-time limit. Nevertheless, implementing time-evolution on a \ac{NISQ} devices is currently impractical, since it requires long coherence times, and a significant number of ancilla qubits and/or precise qubit reset capabilities. One could resort to implementing the Davies generator directly on the quantum computer, however this would once again require applying quantum phase estimation~\cite{Rall2022} or other quantum subroutines related to the quantum Fourier transform~\cite{Chen2023}. Furthermore, Ref.~\cite{Bardet2023} and Ref.~\cite{Kastoryano2016} studied thermalization and Davies generators from the perspective of mathematical physics, showing rapid convergence to the Gibbs state and Gibbs distribution, respectively, under certain conditions. 
Ref.~\cite{Brandao2019} devised a method for preparing the Gibbs state of a local Hamiltonian, assuming that the state has an exponential decay of correlations, and that it satisfies the approximate Markov condition uniformly on non-contractible regions. Ref.~\cite{Zhang2023} also developed a quantum algorithm, based on a local quantum Markov process that can be used to sample from the Gibbs distribution.

Temme et al.~\cite{Temme2011} proposed the quantum Metropolis algorithm, that is capable of preparing Gibbs states through random walks, which is inspired from the Metropolis--Hastings algorithm~\cite{Metropolis1953, Hastings1970}. A quadratic speedup is achieved over the direct implementation of the Metropolis--Hastings algorithm by using Grover's algorithm or related quantum algorithmic techniques~\cite{Poulin2009, Chiang2010}. Using modern quantum algorithmic approaches, further polynomial speedups could be achieved, such as using dimension reduction to sample from Gibbs states~\cite{Bilgin2010}, which requires the use of a precise quantum phase estimation subroutine, among others~\cite{Yung2012, Ozols2012, Chowdhury2017, Wocjan2021}. A similar approach is the one in Ref.~\cite{Riera2012}, which requires the use of a circuit very similar to that used in Shor's algorithm~\cite{Shor1994}, which can only be feasibly implemented on fault-tolerant quantum devices. Ref.~\cite{Franca2018} showed how to adapt “Coupling from the Past”-algorithms proposed by Ref.~\cite{Propp1996}, to sample from the exact Gibbs distribution, rather than an approximation, using the quantum Metropolis algorithm. Nevertheless, compared with the local updates of the classical Metropolis algorithm, all of these algorithms require the implementation of large, global quantum circuits across the whole system, making them infeasible for \ac{NISQ} devices. A recent review of current quantum Gibbs-sampling algorithms can be found in Ref.~\cite{Chen2023}.

A promising approach for \ac{NISQ} devices is \acp{VQA}, where a hybrid quantum--classical approach of minimizing an objective function, using a \ac{PQC} as a variational ansatz, leads to the preparation of the Gibbs state. Some approaches utilize a physically-inspired objective function, the Helmholtz free energy, such in Refs.~\cite{Wu2019, Martyn2019, Chowdhury2020, Foldager2022, Sewell2022, Consiglio2023}. Others employ an engineered cost function~\cite{Premaratne2020, Sagastizabal2021}, while others use \ac{QAOA}-based approaches~\cite{Wu2019, Zhu2020}. Alternative variational approaches consist of using: truncated Taylor series to evaluate an approximation of the free energy~\cite{Wang2021}; \ac{ADAPT}-\ac{VQA} applied to a similar cost function to the free energy~\cite{Warren2022}; and a \ac{VQA} based on McLachlan’s variational principle to initialize and evolve a \ac{TFD} state~\cite{Martyn2019, Wu2019, Zhu2020, Premaratne2020, Sagastizabal2021, Lee2022}. A \ac{TFD} state is defined as a pure state written in basis vectors $\ket{i}_A$ of system $A$, and $\ket{\tilde{i}}_B$ of system $B$:
\begin{equation}
    \ket{\textrm{TFD}(\beta, \cH)} = \sum_{i=0}^{d - 1} \sqrt{\frac{e^{-\beta E_i}}{\cZ(\beta, \cH)}} \ket{i}_A \otimes \ket{\tilde{i}}_B.
    \label{eq:TFD}
\end{equation}
Tracing out system $A$ ($B$) yields the Gibbs state on system $B$ ($A$). \ac{TFD} states also find applications in black hole theory~\cite{Israel1976, Maldacena2003} and teleportation through traversable wormholes~\cite{Gao2017, Maldacena2017, Gao2019}.

Alternative algorithms prepare thermal states through \ac{QITE}, which typically require, either starting from a maximally mixed state~\cite{McArdle2019, Gacon2022, Zoufal2021}, or a maximally entangled state~\cite{Yuan2019}, and carrying out imaginary time evolution of time $\tau$, resulting in preparing the Gibbs state at inverse temperature $\beta$, that is directly proportional to $\tau$. Refs.~\cite{Motta2020, Sun2021, Getelina2023} utilize \ac{METTS}, in combination with \ac{QITE} or measurement-based variational \ac{QITE} to compute thermal averages and correlation functions of Hamiltonians. Ref.~\cite{Tan2020} developed a linear scaling \ac{QITE} algorithm, called Fast \ac{QITE}, that can also compute thermal averages. 
Ref.~\cite{Wang2023} also employed variational \ac{QITE}, that is inspired from the measurement-based approach~\cite{Motta2020}, however, it is adapted to be ansatz-based \ac{QITE}. Ref.~\cite{Shtanko2021} utilizes random quantum circuits with intermediate measurements to impose \ac{QITE}, while Ref.~\cite{Silva2022} proposed a fragmented \ac{QITE} algorithm, for sampling from the Gibbs distribution.
 
Variational ans\"{a}tze based on multi-scale entanglement renormalization~\cite{Sewell2022} and product spectrum ansatz~\cite{Martyn2019} have also been proposed in order to prepare Gibbs states. Other \ac{VQA} approaches rely on variational quantum simulation~\cite{Verdon2019, Guo2023} and variational autoregressive networks~\cite{Liu2021}. The hybrid quantum--classical algorithms proposed in Ref.~\cite{Lu2021} differ from \ac{VQA}, in that they compute microcanonical and canonical properties of many-body systems, using: filtering operators, similar to quantum phase estimation; and Monte Carlo simulations. Ref.~\cite{Cohn2020} employs Green's functions to sample from the Gibbs distribution. Finally, Ref.~\cite{Haug2022} proposed quantum-assisted simulation to prepare thermal states, which, contrary to \acp{VQA}, does not require a hybrid quantum--classical feedback loop.

\subsection{Quantum Boltzmann Machines} \label{sec:QBMs}

One pertinent application of Gibbs states is in \acp{QBM}~\cite{Amin2018}, which are a type of quantum machine learning model that are based on the principles of \acp{CBM}~\cite{Ackley1985}, but incorporate quantum effects such as entanglement and superposition. \acp{CBM} are variants of neural networks embedded in an undirected graph, where the weights and biases of the network represent the information encoded within it. Typically, the network consists of two groups of nodes: the visible nodes, that determine the input and output of the network; and the hidden nodes; which act as latent variables~\cite{Zoufal2021}. The purpose of training a Boltzmann machine is to learn the sets of weights and biases such that the resulting network is capable of approximating a target probability distribution, whereby the model learns by feeding in training data. The Boltzmann machine can then be used for both discriminative and generative learning~\cite{Amin2018}.

A Boltzmann machine is represented by an undirected graph, where $\vV$ is the set of vertices (nodes) and $\vE$ is the set of edges. The states of the \ac{CBM} are defined by $\vZ = \{\vU, \vH\}$, where $\vU$ represents the state of the visible nodes, and $\vH$ represents the state of the hidden nodes. This defines an energy function
\begin{equation}
    E(\vZ) = -\sum_{i \in \vV} b_i z_i - \sum_{\{i, j\} \in \vE} w_{ij} z_i z_j,
    \label{eq:BM_energy}
\end{equation}
with $w_{ij} \in \vw, b_i \in \vb$, denoting the weights and biases (parameters) of the model, respectively. $z_i$ is a binary unit, and to remain consistent with quantum mechanics, we define $z_i \in \{-1, 1\}$.  Thus, the probability to observe a particular configuration $\vU$ of visible nodes is defined as
\begin{equation}
    p_{\vU} = \frac{1}{\cZ(\beta, \vZ)} \sum_{\vH} e^{-\beta E(\vZ)},
    \label{eq:marginal_distribution}
\end{equation}
where in this case,
\begin{equation}
    \cZ(\beta, \vZ) = \sum_{\vZ} e^{-\beta E(\vZ)}.
\end{equation}
The Boltzmann distribution summed over the hidden variables in Eq.~\eqref{eq:marginal_distribution} is called the marginal distribution. The goal of a \ac{CBM} is to determine the weights and biases of \eqref{eq:BM_energy}, such that $p_{\vU}$ approximates well $p^\text{data}_{\vU}$ defined by the training data. Typically, the cost function employed in the optimization procedure of a \ac{CBM} is the Kullback-Leibler divergence, defined as
\begin{equation}
    \cS(p^\text{data}_{\vU} \|  p_{\vU}) = \sum_{\vU} p^\text{data}_{\vU} \log p^\text{data}_{\vU} - \sum_{\vU} p^\text{data}_{\vU} \log p_{\vU}.
\end{equation}

The main difference between a \ac{CBM} and a \ac{QBM}, is that the latter employs nodes that are defined by Pauli matrices rather than binary units, and therefore constructs a network defined by a parameterized Hamiltonian:
\begin{equation}
    \cH(\vb, \vw) = -\sum_{i \in \vV} b_i \sigma^z_i - \sum_{\{i, j\}  \in \vE} w_{ij} \sigma^z_i \sigma^z_j.
    \label{eq:BM_hamiltonian}
\end{equation}
In this case, the \ac{QBM} is defined as the Gibbs state
\begin{equation}
    \rho(\beta, \cH) = \frac{e^{-\beta \cH(\vb, \vw)}}{\cZ(\beta, \cH)},
\end{equation}
where
\begin{equation}
    \cZ(\beta, \cH) = \Tr{e^{-\beta \cH(\vb, \vw)}}.
\end{equation}
The cost function is thus the generalized Kullback-Leibler divergence for density matrices, which corresponds to the quantum relative entropy:
\begin{equation}
    \cS(\sigma \| \rho(\beta, \cH)) = \Tr{\sigma \ln \sigma} - \Tr{\sigma \ln \rho(\beta, \cH)},
\end{equation}
where $\sigma$ is the target density matrix representing the data embedded into a mixed state.

The goal of \ac{QBM} training is to find a set of weights and biases of the Hamiltonian, such that the Gibbs state approximates the target density matrix. There are various methods of tackling the training of \acp{QBM}~\cite{Gacon2022, Zoufal2021, Coopmans2023, Kalis2023, Huijgen2023}, and an analogous problem to \ac{QBM} training is Hamiltonian learning~\cite{Anshu2021}.

\section{Variational Gibbs State Preparation} \label{sec:VGSP}

In this section we will describe how one can prepare a Gibbs state using a \ac{VQA} with the free energy as a cost function, specifically discussing the algorithm presented in Ref.~\cite{Consiglio2023}. Section~\ref{sec:vn_entropy} and~\ref{sec:free_energy} are intended to serve as a qualitative description of the estimation of the von Neumann entropy and the free energy, respectively. Consequently, in Section~\ref{sec:framework}, a quantitative description of the framework of the \ac{VQA} is carried out, mathematically linking the von Neumann entropy and energy expectation, with the estimation of the free energy.

Suppose one chooses a Hamiltonian $\cH$ and inverse temperature $\beta$. For a general state $\rho$, one can define a generalized Helmholtz free energy as
\begin{equation}
    \cF(\rho) = \Tr{\cH \rho} - \beta^{-1}\cS(\rho),
    \label{eq:helmholtz_free_energy}
\end{equation}
where the von Neumann entropy $\cS(\rho)$ can be expressed in terms of the eigenvalues, $p_i$, of $\rho$,
\begin{equation}
    \cS(\rho) = -\sum_{i=0}^{d - 1} p_i \ln p_i.
    \label{eq:von_neumann}
\end{equation}
Since the Gibbs state is the unique state that minimizes the free energy of $\cH$~\cite{Matsui1994}, a variational form can be put forward that takes Eq.~\eqref{eq:helmholtz_free_energy} as an objective function, such that
\begin{equation}
    \rho(\beta, \cH) = \underset{\rho}{\arg\min}~\cF(\rho).
\end{equation}
In this case, $p_i = \exp\left(-\beta E_i\right)/\cZ(\beta, \cH)$ is the probability of getting the eigenstate $\ket{E_i}$ from the ensemble $\rho(\beta, \cH)$.

\subsection{Computing the von Neumann Entropy}~\label{sec:vn_entropy}

The difficulty in measuring the von Neumann entropy, defined by Eq.~\eqref{eq:von_neumann}, of a quantum state on a \ac{NISQ} device is typically the challenging part of variational Gibbs state preparation algorithms, since $\cS(\rho)$ is not an observable. To exactly compute the von Neumann entropy on a quantum device, one would have to perform full state tomography on the Gibbs state, which has a time complexity of $\cO(3^n)$ and a space complexity of $\cO(4^n)$. As a consequence, different works have employed various techniques to compute an approximation for the von Neumann entropy, such as using: a Fourier series approximation to the von Neumann entropy~\cite{Chowdhury2020}; a thermal multi-scale entanglement renormalization ansatz~\cite{Sewell2022}; sums of R\'enyi entropies~\cite{Foldager2022}; among others. On the other hand, Ref.~\cite{Consiglio2023} devised a method of computing the von Neumann exactly, via post-processing of measurement results acting on ancillary register.

When preparing an $n$-qubit state, the unitary gates used in quantum computers ensure that the final quantum state of the entire register, starting from the input state $\ket{0}^{\otimes n}$, remains pure. As a result, in order to prepare an $n$-qubit Gibbs state on the system register, an $m \leq n$-qubit ancillary register is required. For example, in the case of the infinite-temperature Gibbs state, which is a fully mixed state, $m = n$ qubits are needed in the ancillary register to achieve maximal von Neumann entropy. To accurately evaluate the von Neumann entropy without any approximations, the complete Boltzmann distribution is prepared on the ancillary register, thus, $m = n$ is set regardless of the temperature.

\subsection{Computing the Free Energy}~\label{sec:free_energy}

Following the prescription of Ref.~\cite{Consiglio2023}, we shall denote the ancillary register as $A$, while the preparation of the Gibbs state will be carried out on the system register $S$. The purpose of the \ac{VQA} is to effectively create the Boltzmann distribution on $A$, which is then imposed on $S$, via intermediary \textsc{CNOT} gates, to generate a diagonal mixed state in the computational basis. In the ancillary register we can choose a unitary ansatz capable of preparing such a probability distribution. Thus, the ancillary qubits are responsible for classically mixing in the probabilities of the thermal state, while also being able to access these probabilities via measurements in the computational basis. On the other hand, the system register will host the preparation of the Gibbs state, as well as the measurement of the expectation value of our desired Hamiltonian during the optimization routine.

The specific design of the \ac{PQC} allows classical post-processing of simple measurement results, carried out on ancillary qubits in the computational basis, to determine the von Neumann entropy. A diagrammatic representation of the structure of the \ac{PQC} is shown in Fig.~\ref{fig:gibbs_circuit}. Note that while the \ac{PQC} of the algorithm has to have a particular structure --- a unitary acting on the ancillae and a unitary acting on the system, connected by intermediary \textsc{CNOT} gates --- it is not dependent on the choice of Hamiltonian $\cH$, inverse temperature $\beta$, or the variational ans\"atze, $U_A$ and $U_S$, employed within. This is in addition to enjoying a sub-exponential scaling in the number of shots needed to precisely compute the Boltzmann probabilities. A qualitative analysis of a power-law scaling is presented in Section~\ref{sec:error_analysis}.

\subsection{Modular Structure of the PQC}~\label{sec:framework}

\begin{figure}[t]
    \centering
    \includegraphics[width=0.5\textwidth]{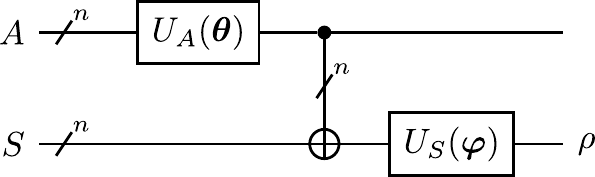}
    \caption{\ac{PQC} for Gibbs state preparation, with systems $A$ and $S$ each carrying $n$ qubits. \textsc{CNOT} gates act between each qubit $A_i$ and corresponding $S_i$.}
    \label{fig:gibbs_circuit}
\end{figure}

The \ac{PQC}, as shown in Fig.~\ref{fig:gibbs_circuit} for the \ac{VQA}, is composed of a unitary gate $U_A$ acting on the ancillary qubits, and a unitary gate $U_S$ acting on the system qubits, with \textsc{CNOT} gates in between. Note that the circuit notation we are using here means that there are $n$ qubits for both the system and the ancillae, as well as $n$ \textsc{CNOT} gates that act in parallel, and are denoted as
\begin{equation}
    \textsc{CNOT}_{AS} \equiv \bigotimes\limits_{i=0}^{n - 1}\textsc{CNOT}_{A_i S_i}.
\end{equation}
The parametrized unitary $U_A$ acting on the ancillae, followed by \textsc{CNOT} gates between the ancillary and system qubits, is responsible for preparing a probability distribution on the system. The parametrized unitary $U_S$ is then applied on the system qubits to transform the computational basis states into the eigenstates of the Hamiltonian.

Throughout the paper we will describe unitary operations and density matrices in the computational basis: $\ket{0} \equiv \left(1~0\right)^\top$ and $\ket{1} \equiv \left(0~1\right)^\top$, which is spanned by one qubit (and products of the basis states in the case of multiple qubits). We will also denote the $n$-fold tensor product of a state $\ket{\psi}$, of an $n$-qubit register $R$, as $\ket{\psi}_R^{\otimes n} \equiv \bigotimes_{i = 0}^{n - 1} \ket{\psi}_i$, where $i \in R$. Denote a general unitary gate of dimension $d = 2^n$, as $U_A = (u_{i, j})_{0\leq i, j \leq d-1}$. Starting with the initial state of the $2n$-qubit register, $\ket{0}_{AS}^{\otimes 2n}$, we apply the unitary gate $U_A$ on the ancillae to get a quantum state $\ket{\psi}_A$, such that
\begin{equation}
    (U_A \otimes I_S) \ket{0}_{AS}^{\otimes 2n} = \ket{\psi}_{A} \otimes \ket{0}_{S}^{\otimes n},
    \label{eq:U_A}
\end{equation}
where $\ket{\psi}_A = \sum_{i=0}^{d - 1} u_{i,0} \ket{i}_A$ and $I_S$ is the identity acting on the system. Since Eq.~\eqref{eq:U_A} is applied to the all-zero state in the ancillary register, then the operation serves to extract the first column of the unitary operator $U_A$. The next step is to prepare a classical probability mixture on the system qubits, which can be done by applying \textsc{CNOT} gates between each ancilla and system qubit. This results in a state
\begin{align}
    \textsc{CNOT}_{AS} \left( \ket{\psi}_{A} \otimes \ket{0}_{S}^{\otimes n} \right) = \sum_{i=0}^{d - 1} u_{i,0} \ket{i}_A \otimes \ket{i}_S.
    \label{eq:pre_tfd}
\end{align}
By then tracing out the ancillary qubits, we arrive at
\begin{align}
    \PTr{A}{\left( \sum_{i=0}^{d-1} u_{i,0}\ket{i}_A \otimes \ket{i}_S \right) \left( \sum_{j=0}^{d-1} u_{j,0}^* \bra{j}_A \otimes \bra{j}_S \right)} = &\sum_{i,j=0}^{d-1} u_{i,0}u_{j,0}^* \braket{i}{j} \ketbra{i}{j}_S \nonumber \\ = \
    &\sum_{i=0}^{d-1} |u_{i,0}|^2 \ketbra{i}{i}_S,
    \label{eq:diag}
\end{align}
ending up with a diagonal mixed state on the system, with probabilities given directly by the absolute square of the entries of the first column of $U_A$, that is, $p_i = |u_{i,0}|^2$. If the system qubits were traced out instead, we would end up with the same diagonal mixed state,
\begin{align}
    \PTr{S}{\left( \sum_{i=0}^{d-1} u_{i,0}\ket{i}_A \otimes \ket{i}_S \right) \left( \sum_{j=0}^{d-1} u_{j,0}^* \bra{j}_A \otimes \bra{j}_S \right)} = &\sum_{i,j=0}^{d-1} u_{i,0}u_{j,0}^* \braket{i}{j} \ketbra{i}{j}_A \nonumber \\ = \
    &\sum_{i=0}^{d-1} |u_{i,0}|^2 \ketbra{i}{i}_A.
\end{align}
This implies that by measuring in the computational basis of the ancillary qubits, we can determine the probabilities $p_i$, which can then be post-processed to determine the von Neumann entropy $\cS$ of the state $\rho$ via Eq.~\eqref{eq:von_neumann} (since the entropy of $A$ is the same as that of $S$). As a result, since $U_A$ only serves to create a probability distribution from the entries of the first column, we can do away with a parametrized orthogonal (real unitary) operator, thus requiring less gates and parameters for the ancillary ansatz.

The unitary gate $U_S$ then serves to transform the computational basis states of the system qubits to the eigenstates of the Gibbs state, once it is optimized by the \ac{VQA}, such that
\begin{align}
    \rho = U_S \left( \sum_{i=0}^{d-1} |u_{i,0}|^2 \ketbra{i}{i}_S \right) U_S^\dagger = \sum_{i=0}^{d - 1} p_i \ketbra{\psi_i},
\end{align}
where the expectation value $\Tr{\cH \rho}$ of the Hamiltonian can be measured. Ideally, at the end of the optimization procedure, $p_i = \exp\left(-\beta E_i\right)/\cZ(\beta, \cH)$ and $\ket{\psi_i} = \ket{E_i}$, so that we get
\begin{equation}
    \rho(\beta, \cH) = \sum_{i=0}^{d - 1} \frac{e^{-\beta E_i}}{\cZ(\beta, \cH)} \ketbra{E_i}.
\end{equation}
The \ac{VQA} therefore avoids the entire difficulty of measuring the von Neumann entropy of a mixed state on a quantum computer, and instead transfers the task of post-processing measurement results to the classical computer, which is much more tractable.

\subsection{Objective Function}

Finally, we can define the objective function of our \ac{VQA} to minimize the free energy~\eqref{eq:helmholtz_free_energy}, via our constructed \ac{PQC}, to obtain the Gibbs state
\begin{align}
    \rho(\beta, \cH) &= \underset{\bm{\theta}, \bm{\varphi}}{\arg\min} \ \cF\left(\rho\left(\bm{\theta}, \bm{\varphi}\right)\right) \nonumber \\ &= \underset{\bm{\theta}, \bm{\varphi}}{\arg\min} \left( \Tr{\cH \rho_S(\bm{\theta}, \bm{\varphi})} - \beta^{-1}\cS\left(\rho_A(\bm{\theta})\right) \right).
    \label{eq:gibbs_state}
\end{align}

It is noteworthy to mention that while the energy expectation depends on both sets of angles $\bm{\theta}$ (as $U_A$ is responsible for parameterizing the Boltzmann distribution) and $\bm{\varphi}$ (as $U_S$ is responsible for parameterizing the eigenstates of the Gibbs state), the calculation of the von Neumann entropy only depends on $\bm{\theta}$.

\begin{figure}[t]
    \centering
    \includegraphics[width=0.6\textwidth]{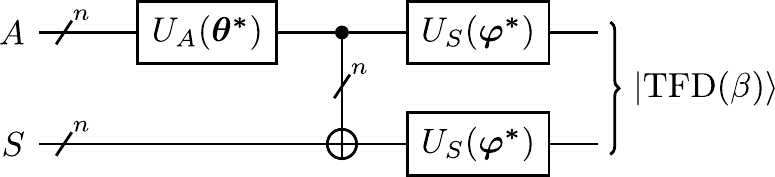}
    \caption{Optimal \ac{PQC} for \ac{TFD} state preparation, with systems $A$ and $S$ each carrying $n$ qubits. \textsc{CNOT} gates act between each qubit $A_i$ and corresponding $S_i$.}
    \label{fig:tfd_circuit}
\end{figure}

Furthermore, once we obtain the optimal parameters $\bm{\theta^*}$, $\bm{\varphi^*}$, preparing the Gibbs state $\rho(\beta, \cH)$ on the system qubits $S$, one can place the same unitary $U_S$ with optimal parameters $\bm{\varphi^*}$ on the ancillary qubits to prepare the \ac{TFD} state on the quantum computer, as shown in Fig.~\ref{fig:tfd_circuit}. Notice that this is equivalent to Eq.~\eqref{eq:pre_tfd} followed by applying the optimized $U_S$ on both registers. Tracing out either the ancilla or system register yields the same Gibbs state on the other register.

\section{Preparing Gibbs states of the \textit{XY} model} \label{sec:results}

In this Section we assess the performance of the \ac{VQA} for Gibbs state preparation of an $XY$ model. The $XY$ model~\cite{Lieb1961} is defined as
\begin{equation}
    \cH = -\sum_{i=1}^n \left( \frac{1 + \gamma}{2} \sigma_i^x \sigma_{i+1}^x + \frac{1 - \gamma}{2} \sigma_i^y \sigma_{i+1}^y \right) - h \sum_{i=1}^n \sigma_i^z.
    \label{eq_XY}
\end{equation}
Here we only report one relevant property for implementing a problem-inspired ansatz for $U_S$. The Hamiltonian in Eq.~\eqref{eq_XY} commutes with the parity operator $\cP=\prod_{i=0}^{n - 1} \sigma^z_i$. As a consequence, the eigenstates of $\cH$ have definite parity, and so will the eigenstates of $\rho_{\beta}$.

To assess the performance of the \ac{VQA}, we utilize the Uhlmann-Josza fidelity as a figure of merit~\cite{Uhlmann2011}, defined as $F\left(\rho,\sigma\right) = \left(\Tr{\sqrt{\sqrt{\rho}\sigma\sqrt{\rho}}}\right)^2$. This fidelity measure quantifies the ``closeness'' between the prepared state and the target Gibbs state, making it a commonly-employed metric for distinguishability. However, alternative measures exist, each with its own interpretation. For instance, the trace distance~\cite{Nielsen2010} can be used, which guarantees that if its value between two states is bounded by $\epsilon$, the expectation values computed on the effectively prepared state will differ from those of the true Gibbs state by at most an amount proportional to $\epsilon$\cite{Holmes2022}. Another option is the relative entropy\cite{Nielsen2010}, which characterizes the distinguishability between the two states as the surprise that arises when an event occurs that is not possible with the true Gibbs state~\cite{Vedral2002}.

We use an alternating, entangling brick-wall \ac{PQC} for the unitary $U_A$, composed of parametrized $R_Y(\theta_i)$ gates, and \textsc{CNOT}s as the entangling gates. This ansatz is hardware efficient and is sufficient to produce real amplitudes for preparing the probability distribution. Note that we require the use of entangling gates~\cite{Consiglio2023}, as otherwise we will not be able to prepare any arbitrary probability distribution, including the Boltzmann distribution of the $XY$ model.

\begin{figure}[t]
    \centering
    \includegraphics[width=\textwidth]{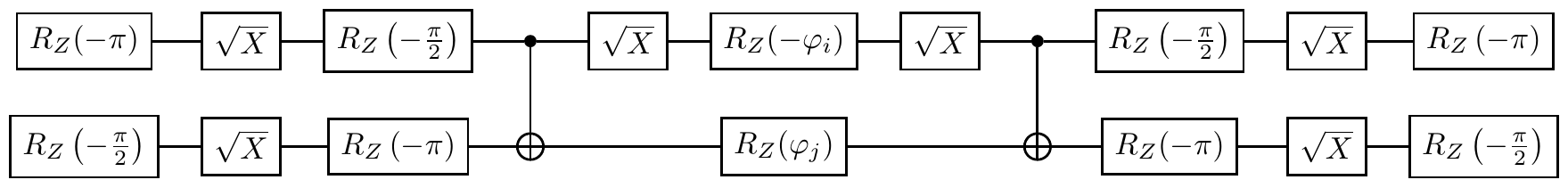}
    \caption{Decomposed $R_P$ gate in Eq.~\eqref{eq:R_P}.}
    \label{fig:R_P}
\end{figure}

\begin{figure}[t]
    \centering
    \includegraphics[width=\textwidth]{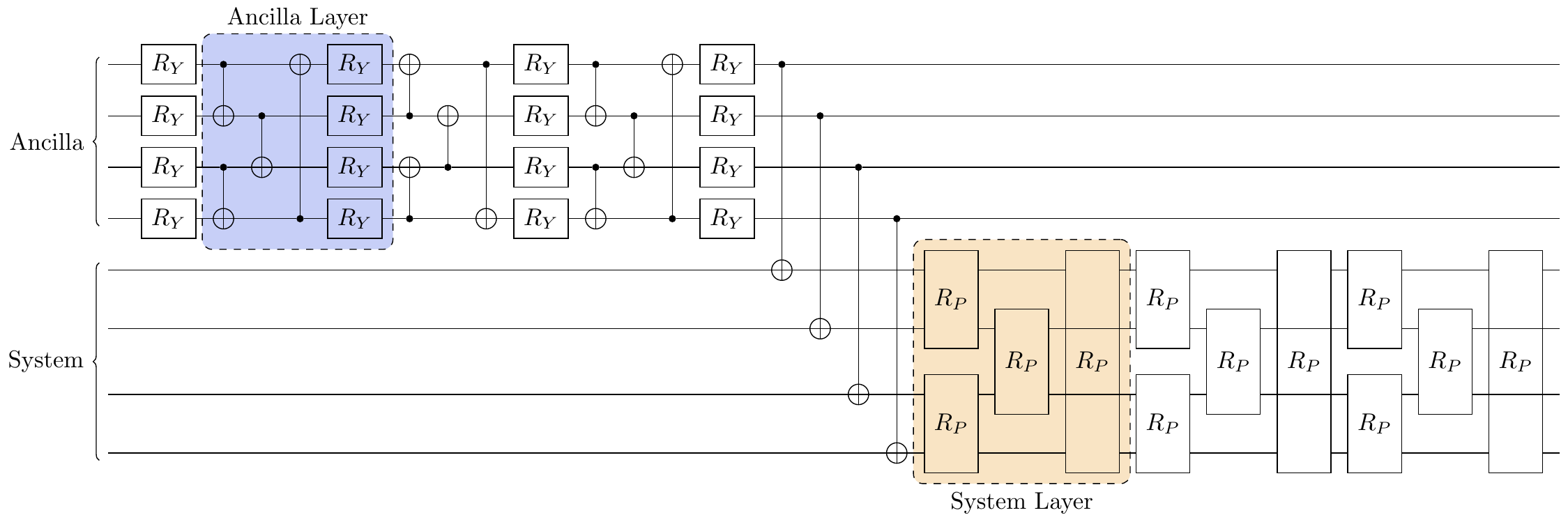}
    \caption{Example of an eight-qubit \ac{PQC}, consisting of three ancilla layers acting on a four-qubit register, and three system layers acting on another four-qubit register. Each $R_Y$ gate is parametrized with one parameter $\theta_i$, while each $R_P$ gate has two parameters $\varphi_i$ and $\varphi_j$. The $R_P$ gate is defined in Eq.~\eqref{eq:R_P} with its decomposition shown in Fig.~\ref{fig:R_P}. Note that the intermediary \textsc{CNOT} gates, as well as the \textsc{CNOT} and $R_P$ gates acting on qubits two and three, and on qubits one and four of the ancilla and system, respectively, can be carried out in parallel, respectively.}
    \label{fig:pqc_4qubits}
\end{figure}

For the unitary $U_S$, We employ a brick-wall structure solely using parity-preserving gates --- $R_{XY}(\varphi_i)$ followed by $R_{YX}(\varphi_j)$ gates. If we combine the two gates, which we denote as $R_P(\varphi_i, \varphi_j)$, we get
\begin{align}
    R_P(\varphi_i, \varphi_j) &= R_{YX}(\varphi_j)\cdot R_{XY}(\varphi_i) \nonumber \\
    &=
    \left(
    \begin{array}{cccc}
     \cos \left(\frac{\varphi_i +\varphi_j }{2}\right) & 0 & 0 & \sin \left(\frac{\varphi_i +\varphi_j }{2}\right) \\
     0 & \cos \left(\frac{\varphi_i -\varphi_j }{2}\right) & -\sin \left(\frac{\varphi_i -\varphi_j }{2}\right) & 0 \\
     0 & \sin \left(\frac{\varphi_i -\varphi_j }{2}\right) & \cos \left(\frac{\varphi_i -\varphi_j }{2}\right) & 0 \\
     -\sin \left(\frac{\varphi_i +\varphi_j }{2}\right) & 0 & 0 & \cos \left(\frac{\varphi_i +\varphi_j }{2}\right) \\
    \end{array}
    \right),
    \label{eq:R_P}
\end{align}
which can be decomposed into two \textsc{CNOT} gates six $\sqrt{X}$ gates, and ten $R_Z$ gates. The decomposed unitary is shown in Fig.~\ref{fig:R_P}. One layer of the unitary acting on the system qubits consists of a brick-wall structure, composed of an even-odd sublayer of $R_P$ gates, followed by an odd-even sublayer of $R_P$ gates. Table~\ref{tab:scaling} shows the scaling of the \ac{VQA} assuming a closed ladder connectivity, for $n > 2$. An example of a \ac{PQC}, split into a four-qubit ancilla register, and a four-qubit system register, is shown in Fig.~\ref{fig:pqc_4qubits}.

\begin{table}[t]
\caption{Scaling of the \ac{VQA} assuming a closed ladder connectivity, for $n > 2$, where $l_A$ and $l_S$ are the number of ancilla ansatz and system ansatz layers, respectively, and $P$ is 2 when $n$ is even and 3 when $n$ is odd. Circuit depth is calculated on the depth of \textsc{CNOT} gates.}
\centering
\begin{tabular}{|l|l|l|} 
    \hline
    \# of parameters & $n(l_A + 1) + 2nl_S$ & $\cO(n(l_A + l_S))$ \\ 
    \hline
    \# of \textsc{CNOT} gates & $nl_A + 2nl_S + n$ & $\cO(n(l_A + l_S))$ \\
    \hline
    \# of $\sqrt{X}$ gates & $2n(l_A + 1) + 6nl_S$ & $\cO(n(l_A + l_S))$ \\ 
    \hline
    Circuit depth & $P l_A + 2Pl_S + 1$ & $\cO(l_A + l_S)$ \\ 
    \hline
\end{tabular}
\label{tab:scaling}
\end{table}

\subsection{Statevector Results} \label{sec:statevector_results}

\begin{figure}[!ht]
    \centering
    \includegraphics[width=\textwidth]{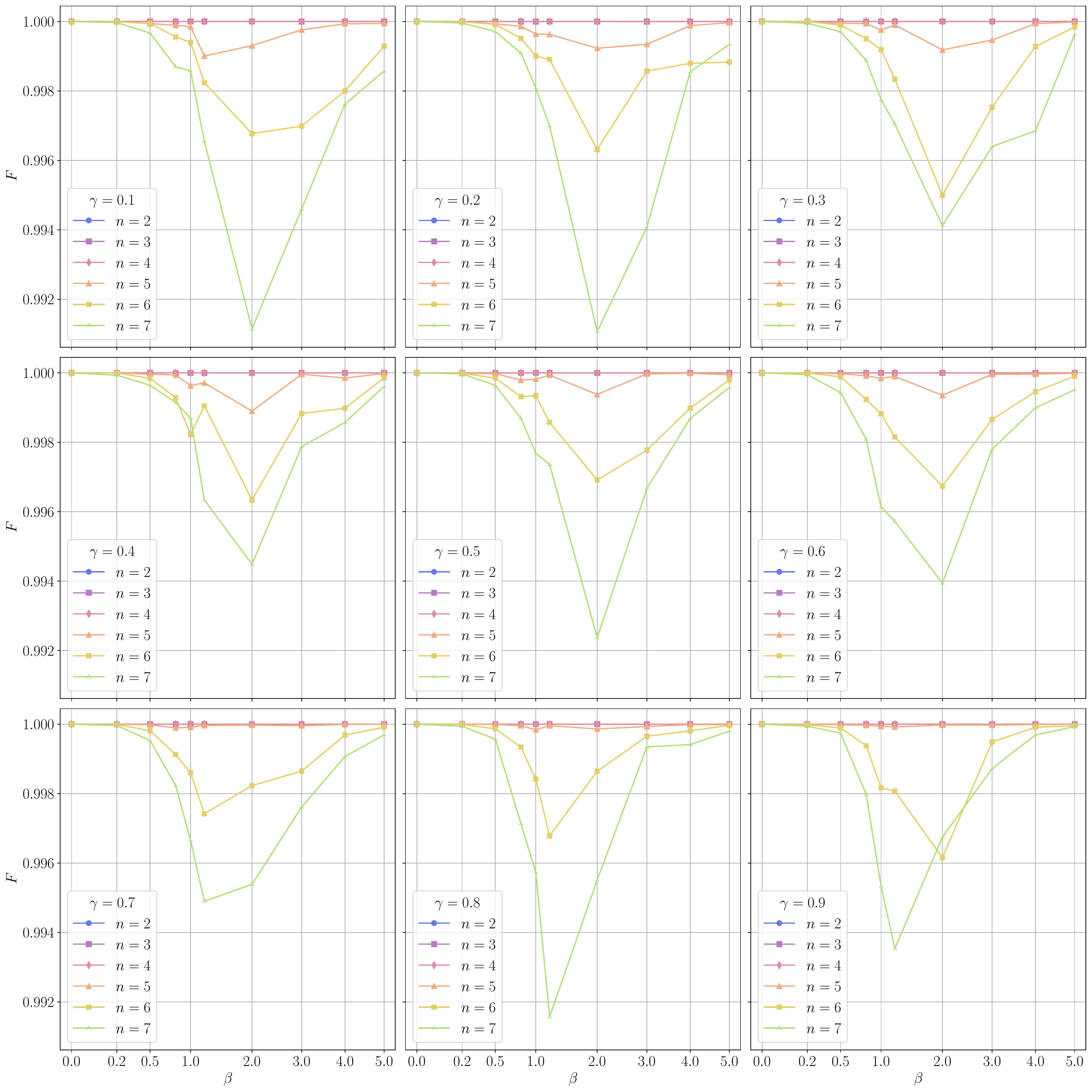}
    \caption{Fidelity $F$, of the obtained state via statevector simulations (using \ac{BFGS}) with the exact Gibbs state, vs inverse temperature $\beta$, for two to seven qubits of the $XY$ model with $\gamma$ between 0.1 and 0.9, and $h = 0.5$. A total of ten runs are made for each point, with the optimal state taken to be the one that maximizes the fidelity.}
    \label{fig:statevector_fidelity}
\end{figure}

Fig.~\ref{fig:statevector_fidelity} shows the fidelity of the generated mixed state when compared with the exact Gibbs state of the $XY$ model with $h = 0.5$, and $\gamma$ ranging from 0.1 to 0.9 in steps of 0.1, across a range of temperatures for system sizes between two to seven qubits. The \ac{VQA} was carried out using statevector simulations with the \ac{BFGS} optimizer~\cite{Nocedal2006}. We used $n - 1$ layers for both the ancilla ansatz and for the system ansatz, with the scaling highlighted in Table~\ref{tab:scaling_2}. The number of layers was heuristically chosen to satisfy, at most, a polynomial scaling in quantum resources (but linear in depth), while achieving a fidelity higher than 99\%. Furthermore, in order to alleviate the issue of getting stuck in local minima, the optimizer was embedded in a Monte Carlo framework, that is, taking ten random initial positions and carrying out a local optimization from each position --- which we call a `run' --- and finally taking the global minimum to be the minimum over all runs.

\begin{table}[t]
\caption{Scaling of the \ac{VQA} assuming a closed ladder connectivity, for $n > 2$, with $l_A = n - 1$, and $l_S = n - 1$, and $P$ is 2 when $n$ is even and 3 when $n$ is odd. Circuit depth is calculated on the depth of \textsc{CNOT} gates.}
\centering
\begin{tabular}{|l|l|l|} 
    \hline
    \# of parameters & $n(3n - 2)$ & $\cO(n^2)$ \\ 
    \hline
    \# of \textsc{CNOT} gates & $3n^2 - 2n$ & $\cO(n^2)$ \\
    \hline
    \# of $\sqrt{X}$ gates & $2n(4n - 3)$ & $\cO(n^2)$ \\ 
    \hline
    Circuit depth & $3P(n - 1) + 1$ & $\cO(n)$ \\
    \hline
\end{tabular}
\label{tab:scaling_2}
\end{table}

A total of ten runs of \ac{BFGS} per $\beta$ were carried out to verify the reachability of the \ac{PQC}, with Fig.~\ref{fig:statevector_fidelity} showcasing the maximal fidelity achieved for each $\beta$ out of all runs. The results show that, indeed, our \ac{VQA}, is able to reach a very high fidelity $F > 99\%$ for up to seven-qubit Gibbs states of the $XY$ model. In the case of the extremal points, that is $\beta \rightarrow 0$ and $\beta \rightarrow \infty$, the fidelity reaches unity, for all investigated system sizes. For $\beta \rightarrow 0$, the problem reduces to maximizing the von Neumann entropy, where the problem becomes independent of the energy, and therefore, independent of the prepared eigenstates by $U_S$. For $\beta \rightarrow \infty$, the problem reduces to minimizing the energy, which is equivalent to finding the ground-state of the Hamiltonian, and so, independent of $U_A$. This is in contrast with intermediary temperatures $\beta \sim 1$, where the fidelity decreases with the number of qubits. This could be attributed to the fact that at intermediate temperatures, a substantial amount of Boltzmann probabilities are necessary for evaluating the von Neumann entropy with a high precision, and similarly, a large number of eigenstates contribute to preparing the Gibbs state, resulting in a dip in the fidelity at those temperatures.

\subsection{Error Analysis}\label{sec:error_analysis}

\begin{figure}[!ht]
    \centering
    \includegraphics[width=\textwidth]{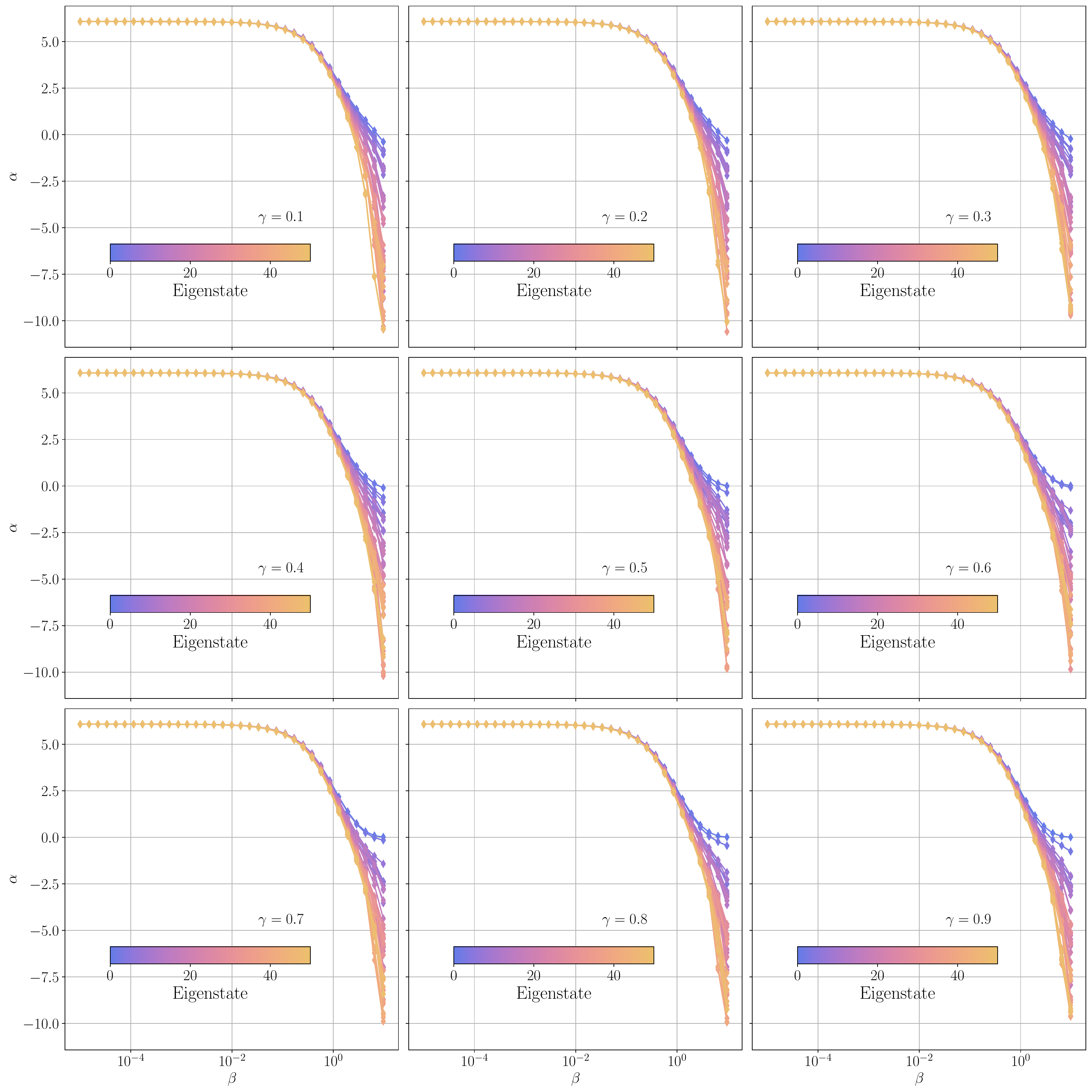}
    \caption{The exponent $\alpha_i$ in Eq.~\eqref{eq:fit} is plotted vs $\beta$ for the first 51 $p_i$ of the $XY$ model with $\gamma$ between 0.1 and 0.9, and $h = 0.5$, where each point is the polynomial fit for $n$ between 8 and 20.}
    \label{fig:shots_scaling}
\end{figure}

In this Section, we perform an analysis of the error scaling with the number of shots, in faithfully identifying the probability distribution, $\bm{p} = \{p_0, p_1, \dots, p_{d-1}\}$, where $d = 2^n$, prepared by $U_A$. The outcome of one shot of a quantum circuit can be described by a multinomial distribution, where $p_i$ is the probability of observing a bit string $i$. The expected value of a multinomially distributed random bit string $i$ is
\begin{equation}
    \mu = \mathbb{E}[i] = N_s p_i,
\end{equation}
with the variance being
\begin{equation}
    \sigma^2 = \mathrm{Var}[i] = N_s p_i(1 - p_i),
\end{equation}
where $N_s$ is the number of shots. A quantity that can describe the precision of measuring the bit string $i$, $p_i$ times, is the coefficient of variation, or relative standard deviation,
\begin{equation}
c_\text{v} = \frac{\sigma}{\mu} = \sqrt{\frac{1-p_i}{N_s p_i}}.
\end{equation}
Given that $p_i = \exp(-\beta E_i) / \cZ(\beta, \cH)$ for the Boltzmann distribution, we get
\begin{equation}
    c_\text{v} = \sqrt{\frac{1-\frac{e^{-\beta E_i}}{\cZ(\beta, \cH)}}{N_\text{s}\frac{e^{-\beta E_i}}{\cZ(\beta, \cH)}}} = \sqrt{\frac{\cZ(\beta, \cH) - e^{-\beta E_i}}{N_\text{s}e^{-\beta E_i}}} = \sqrt{\frac{1}{N_\text{s}}\left( \frac{\cZ(\beta, \cH)}{e^{-\beta E_i}} - 1 \right)}.
    \label{eq:c_v}
\end{equation}
From Eq.~\eqref{eq:c_v}, if $\beta \rightarrow 0$, then $\cZ(\beta, \cH) / \exp(-\beta E_i) \rightarrow d~\forall i \in [d]$, which implies
\begin{equation}
    c_\text{v}(\beta \rightarrow 0) = \sqrt{\frac{d - 1}{N_\text{s}}},
\end{equation}
hence, exhibiting an exponential scaling in the number of shots needed for preparing the flat distribution. On the other hand, if $\beta \rightarrow \infty$, then $\cZ(\beta, \cH) / \exp(-\beta E_0) \rightarrow 1$, while all other probabilities tend to zero, and the algorithm reduces to finding the ground-state of the Hamiltonian. As a result,
\begin{equation}
    c_\text{v}(\beta \rightarrow \infty) = 0.
    \label{eq:infinite_temperature}
\end{equation}
For intermediary temperatures, we carry out a qualitative analysis. Fig.~\ref{fig:shots_scaling} reports the exponent $\alpha_i$, of a polynomial fit of the normalized coefficient of variation, defined as 
\begin{equation}
    c_\text{v} \sqrt{N_\text{s}} = C n^{\alpha_i},
    \label{eq:fit}
\end{equation}
for $n$ between 8 and 20, at different $\beta$, where the first 51 $p_i$ of the $XY$ model with $h = 0.5$, and $\gamma$ between 0.1 and 0.9 are considered. We observe that as $\beta \rightarrow \infty$, the exponent $\alpha_0$ approaches zero only for the ground-state occupation probability $p_0$, in accordance with Eq.~\eqref{eq:infinite_temperature}. Simultaneously, as the occupation probability $p_i$ of an excited state decreases to negligible values, the corresponding exponent $\alpha_i$ becomes increasingly negative. However, for any finite temperature, we find that a polynomial fit, with $\alpha_i \lesssim 6$, provides a good approximation of the normalized coefficient of variation. This suggests that achieving a faithful reconstruction of the probability distribution, for the investigated system sizes of the $XY$ model in \ac{NISQ} algorithms, requires a sub-exponential scaling in the number of shots.

\section{Conclusion} \label{sec:conclusion}

In this manuscript, we provided a concise overview of various Gibbs state preparation and Gibbs-sampling algorithms. However, our primary focus was on investigating \acp{VQA} for Gibbs state preparation. To accomplish this, we conducted a benchmark study using one of the latest variational Gibbs state preparation algorithms on the $XY$ model, considering a wide range of temperatures and $\gamma$ coefficients. Following the approach proposed by Consiglio et al.~\cite{Consiglio2023}, we leveraged the unique property of the Gibbs state as the state that minimizes the Helmholtz free energy, which serves as an appropriate objective function for the \ac{VQA}. Through extensive statevector simulations, we achieved fidelities of $F > 99\%$ for system sizes up to seven qubits.

Furthermore, we performed a qualitative analysis of the scalability of the VQA for implementation on a \ac{NISQ} device. We found that the number of gates, iterations, and shots scale at most polynomially with the number of qubits in the $XY$ model. The algorithm's scalability not only makes it suitable for near-term applications, but it holds significant potential for advancing quantum thermodynamic experiments on quantum computers, and ensuring the faithful preparation of Gibbs states for various computational tasks.

The Python code for running the statevector simulations, using Qulacs~\cite{Suzuki2021}, can be found on GitHub~\cite{github}.

\section*{Acknowledgments}

MC would like to thank Jacopo Settino, Andrea Giordano, Carlo Mastroianni, Francesco Plastina, Salvatore Lorenzo, Sabrina Maniscalco, John Goold, and Tony J. G. Apollaro for fruitful and extensive discussions on Gibbs states, and the challenging task of variationally preparing them on \ac{NISQ} computers. MC acknowledges funding by TESS (Tertiary Education Scholarships Scheme), and project QVAQT (Quantum Variational Algorithms for Quantum Technologies) REP-2022-003 financed by the Malta Council for Science \& Technology, for and on behalf of the Foundation for Science and Technology, through the FUSION: R\&I Research Excellence Programme.

\bibliographystyle{splncs04}
\bibliography{ref.bib}

\appendix
\acresetall

\end{document}